\documentclass[preprint,prd,amsmath,amssymb,amsthm,nofootinbib]{revtex4}
\usepackage[mathscr]{euscript}
\usepackage{threeparttable}
\usepackage{subfigure}
\usepackage{bm}
\usepackage{textcomp}
\usepackage{graphicx}
\usepackage{subfigure}
\usepackage{overpic}
\usepackage{multirow}
\usepackage{floatrow}
\usepackage{float} 
\usepackage{amsmath,amssymb,amsthm}
\usepackage{cases}
\usepackage[colorlinks=true,linkcolor=red]{hyperref}
\usepackage[normalem]{ulem}
\newfloatcommand{capbtabbox}{table}[][\FBwidth]

\begin{document}

	\title{Revisiting cosmic acceleration with DESI BAO }
	
	\author{ Jincheng Wang\footnote{J.C.Wang@hunnu.edu.cn},   Hongwei Yu\footnote{hwyu@hunnu.edu.cn} and Puxun Wu\footnote{ pxwu@hunnu.edu.cn} }
	\affiliation{Department of Physics, Institute of Interdisciplinary Studies and Synergetic Innovation Center for Quantum Effects and Applications, Hunan Normal University, Changsha, Hunan 410081, China }
	
	\begin{abstract}
We revisit the evolution of cosmic acceleration in a spatially flat $w_0w_a$CDM universe,  in which the equation of state of dark energy takes the CPL parametrization, using the latest baryon acoustic oscillation (BAO) measurements from the Dark Energy Spectroscopic Instrument (DESI), in combination with Planck cosmic microwave background (CMB) data and several type Ia supernova datasets, including PantheonPlus, Union3, and DESY5. We analyze the deceleration parameter $q(z)$ and the jerk parameter $j(z)$ and further validate our results using the $Om(z)$ diagnostic. Our findings indicate significant deviations from the predictions of the $\Lambda$CDM model. Specifically, DESI BAO, DESI BAO + CMB, DESI BAO + CMB + Union3, and DESI BAO + CMB + DESY5 all provide strong evidence for a slowing down of cosmic acceleration at late times, as indicated by $j(0) < 0$ at more than 1$\sigma$ confidence level,   within the framework of $w_0w_a$CDM  model. These results suggest that  in the $w_0w_a$CDM  universe  cosmic acceleration has already peaked and is now in a phase of decline.
	\end{abstract}

	
	\maketitle
	\section{Introduction}
	\label{sec_in}
In 1998, two independent type Ia supernova (SN Ia) groups discovered that the universe is expanding at an accelerating rate~\cite{riess1998observational,perlmutter1999measurements}. This groundbreaking discovery has since been confirmed by multiple observations, including baryon acoustic oscillations (BAO)\cite{Daniel2005} and cosmic microwave background radiation (CMB)\cite{hinshaw2013nine,aghanim2020planck}. To explain the observed acceleration, dark energy is typically introduced as an additional energy component of the universe. The simplest and most widely accepted candidate for dark energy is the cosmological constant, $\Lambda$, which has an equation of state exactly equal to $-1$. The standard $\Lambda$CDM model, which includes both $\Lambda$ and cold dark matter, provides an excellent fit to various observational datasets and predicts that cosmic acceleration should continue to increase over time as dark energy dominates the universe's energy budget.

However, about fifteen years ago, Shafieloo et al.~\cite{shafieloo2009cosmic} analyzed cosmic expansion using observational data and found indications that cosmic acceleration may have already peaked and is now slowing down. Their analysis was based on a combination of the Constitution SN Ia dataset\cite{hicken2009improved,hicken2009cfa3} and BAO distance measurements at redshifts $z=0.2$ and $z=0.35$ from the 2dFGRS and SDSS surveys~\cite{percival2010baryon}. However, when they incorporated the CMB shift parameter from WMAP~\cite{WMAP:2008lyn}, the evidence for a decline in acceleration disappeared, and the results aligned with the predictions of $\Lambda$CDM. This sparked widespread interest in whether the apparent slowdown of cosmic acceleration was a real physical phenomenon or a transient feature of the data~\cite{Fabris2010,Gong2010,Guimaraes2011,Li2010,Li2011,Li20112,Cai2011,Bolotin2012,Vargas2012,Cardenas2012,Lin2013,Magana2014,Shahalam2015,Bernal2017,Zhang2018,Velten2018,Haridasu2018,Fortunato2024,Wang2025}. Since different conclusions have been drawn from different datasets, new observational data are essential to further investigate the nature of cosmic acceleration. 

Recently, the Dark Energy Spectroscopic Instrument (DESI)~\cite{Adame2024a} released its first-year BAO measurements~\cite{Adame2024desi} based on galaxy, quasar, and Lyman-$\alpha$ forest tracers. The DESI BAO dataset consists of 12 data points, measuring the transverse comoving distance and Hubble rate, or their combination, relative to the sound horizon across seven redshift bins~\cite{Adame2024desi}. While DESI BAO alone is well-fitted by the standard $\Lambda$CDM model, when combined with Planck CMB~\cite{aghanim2020planck} and SN Ia datasets, including PantheonPlus~\cite{brout2022PantheonPlus}, Union3~\cite{rubin2023union}, and DESY5~\cite{abbott2024dark}, a preference for dynamical dark energy emerges at more than 2$\sigma$ confidence level (CL). Consequently, DESI BAO has been widely used to probe dark energy properties and the cosmic expansion history~\cite{Calderon2024,Raamsdonk2024,Tada2024,Yin2024,Luongo2024,Cortes2024,DWang2024,Colgain2024a,Berghaus2024,Giare2024a,Wang20242,Yang2024,Park2024,Wang20243,Huang2024,Dinda2024,Bousis2024,Lodha2025,Ramadan2024,Roy2024,Wang20244,Gialamas2025,Chudaykin2024,Orchard2024,Li2024,Du2024,Giare2024,Dinda2025,Pang2024,David2024,Wolf2024,Almada2024,Li20242,Colgain2024,Zheng2024,Gao2024,Notari2024,Odintsov2024,Odintsov2025a,Odintsov2025b,Marcus2025,Mishra2025,Sousa2025,Huang2025,Li2025,Giare2025,Efstathiou2025,Carloni2025a,Notari2025,Valent2025,Liu2025, Pang2025,Chatrchyan2025,Wolf2025,Shlivko2025,LiTN2025,Taule2025,Qu2025,Tang2025,Carloni2025,Kessler2025,Chunyu2025,Chakraborty2025,Wang2025a,Braglia2025,Eoin2025,Gialamas2025a}. More recently, DESI BAO has been used to reconstruct the evolution of cosmological parameters such as the Hubble parameter $H(z)$ and the deceleration parameter $q(z)$ using the crossing statistics method~\cite{Shafieloo2012,Calderon2024}. This analysis suggests that a universe with $q(0) > 0$, following an accelerating phase ($q(z) < 0$), remains consistent with current data, opening the door to models where cosmic acceleration is slowing down.

In this paper, we revisit the evolution of cosmic acceleration using the latest DESI BAO measurements~\cite{Adame2024desi} in combination with Planck CMB~\cite{aghanim2020planck} and SN Ia datasets from PantheonPlus~\cite{brout2022PantheonPlus}, Union3~\cite{rubin2023union}, and DESY5~\cite{abbott2024dark}. In addition to analyzing the evolution of the deceleration parameter $q(z)$, we explore the jerk parameter $j(z)$, which characterizes variations in $q(z)$ during cosmic expansion. The paper is organized as follows: Section~\ref{sec2} describes the methodology and datasets used. Section~\ref{sec3} presents our results and discussion, while Section~\ref{sec4} summarizes our conclusions.

	\section{Method  and Data}
	\label{sec2}
	\subsection{Method}
	To describe the evolutionary  behavior  of the universe, we introduce the deceleration parameter $q(z)$ defined as
	\begin{align}
		q \equiv - \frac{a \ddot{a}}{\dot{a}^2},
	\end{align}
where  $a=1/(1+z)$ is the cosmic scale factor   and an overdot denotes  differentiation  with respect to cosmic time $t$. A negative value of $q$ indicates  an accelerating expansion, while a positive value signifies a decelerating universe.
To further examine the variation of cosmic acceleration, we consider the jerk parameter $j(z)$~\cite{alam2003exploring,visser2004jerk,rapetti2007kinematical,blandford2004cosmokinetics,sahni2003statefinder}, given by
	\begin{align}
		j \equiv \frac{ \dddot{a}}{a H^2}. 
	\end{align}
The jerk parameter  provides higher-order insights into cosmic expansion. Here $H=\dot{a}/a$ is the Hubble parameter. In the flat $ \Lambda $CDM model, the jerk parameter  is a constant equal to  one. Any deviation of  of $j(z)$  from unity implies that dark energy  is not a cosmological constant but may not indicate that it is dynamically evolving.  Specifically,  $ j(z) > 0 $ means an increasing acceleration, whereas $ j(z) < 0 $ implies a slowing down of cosmic acceleration.   
 In order to  further check the results inferred  from the deceleration and jerk parameters, we also study the $Om$ diagnostic~\cite{sahni2008two, zunckel2008consistency}, defined as: 
	\begin{align} 
		Om(z) \equiv \frac{h^2(z) - 1}{(1+z)^3 - 1}, \quad h(z) = \frac{H(z)}{H_0}.
	\end{align}
Here, $H_0=H(0)$ is the Hubble constant. 	The $Om(z)$ parameter provides a simple yet powerful tool for distinguishing dark energy models. 
	In the flat $\Lambda$CDM model, $Om(z) = \Omega_{m0}$ is constant, where $\Omega_{m0}$ is the present dimensionless matter density parameter.
At low redshifts ($z \ll 1$), since $w \approx [Om(z)-1]/(1-\Omega_{m0})$ with $w$ being the equation of state for dark energy,	 a higher $Om(z)$ indicates a larger $w$.    Moreover, a redshift-dependent evolution of $Om(z)$ in the low-redshift regime indicates a dynamically evolving $w(z)$, and hence signals the presence of dynamical dark energy.  By jointly analyzing the evolution of $ q(z)$,   $ j(z) $ and $ Om(z) $,  we can identify key transitions in cosmic expansion and probe potential deviations from the standard $\Lambda$CDM model.
 
Before investigating  the evolution of these parameters, we first need to know the form of the Hubble parameter. 
Recent studies have shown that dark energy reconstructed using crossing statistics is consistent with results from the Chevallier-Polarski-Linder (CPL) equation of state parametrization~\cite{Calderon2024}. Therefore, we adopt the CPL parameterization to describe the Hubble parameter.  The CPL parametrization~\cite{chevallier2001accelerating,linder2003exploring} models the dark energy equation of state (EoS) as:
	\begin{align}
		w(z) = w_0 + w_a \frac{z}{1+z},
	\end{align}
	where $ w_0 $ represents the present-day the equation of state parameter and $ w_a $ characterizes its evolution. For the cosmological model consisting of dark energy with the CPL parametrization and  cold dark matter ($w_0w_a$CDM), its Hubble parameter can be expressed as 
        \begin{align}
		H^2(z) = H_0^2 \left[ \Omega_{m0} (1+z)^3 + (1-\Omega_{m0}) \exp \left( 3 (1+w_0+w_a)\ln (1+z)-3w_a \frac {z}{1+z} \right) \right].
	\end{align}

	\subsection{Data}

	In order to determine the evolution  of parameters: $Om(z)$, $ q(z) $ and $j(z)$, it is  first necessary to constrain model parameters: $ \omega_0 $, $ \omega_a $, and $ \Omega_{m0} $.
	We employ the Markov Chain Monte Carlo (MCMC) package \texttt{Cobaya}~\cite{torrado2019cobaya,torrado2021cobaya} and Code for Anisotropies in the Microwave Background (CAMB)~\cite{lewis2000efficient,howlett2012cmb} to infer the posterior distributions of these model parameters.  We adopt flat prior distributions for $w_0 \in [-3,1]$ and $w_a \in [-3,2]$, with the additional condition $w_0 + w_a < 0 $  to ensure a phase of matter domination at high redshifts. We assess the convergence of the MCMC chains using the Gelman-Rubin statistic, with $ R-1 < 0.02 $ ~\cite{gelman1992inference}. The chains are analyzed using the publicly available package \texttt{Getdist} \cite{lewis2019getdist}. 
	
	The observational datasets used  in our analysis  are as follows:
	\begin{itemize}
	
		\item \textbf{DESI BAO}: The DESI has provided precise BAO measurements~\cite{Adame2024desi} through the analysis of spatial distributions of galaxies, quasars~\cite{adame2024desib}, and Lyman-$\alpha$~\cite{adame2024desic} forest tracers. These observations span a wide redshift range from $ z \sim 0.1 $ to $ z \sim 4.2 $. The extracted BAO signals enable the measurement of key cosmological quantities, such as the transverse comoving distance, the Hubble horizon, and the angle-averaged distance. Based on the data from the first year of DESI observations, these results have been systematically analyzed and summarized in Tab. 1 in~\cite{Adame2024desi}, offering robust constraints on the expansion rate of the universe and critical parameters governing the dynamics of dark energy.

		\item \textbf{CMB}: The analysis incorporates \texttt{Planck 2018} data~\cite{aghanim2020planck}, covering both temperature and polarization measurements across a wide range of angular scales. Specifically, the \texttt{planck\_2018\_lowl.TT} and \texttt{planck\_2018\_lowl.EE} likelihoods are employed to include large-scale ($ 2 \leq \ell \leq 30 $) temperature (TT) and polarization (EE) power spectra, processed using the \texttt{Commander} and \texttt{SimAll} methods. Additionally, high-resolution ($ \ell > 30 $) CMB data are incorporated via the \texttt{planck\_NPIPE\_highl\_CamSpec.TTTEEE} likelihood, which includes temperature (TT), polarization (EE), and temperature-polarization cross-spectra (TE) data, processed using the \texttt{plik} likelihood method. Finally, we use the \texttt{planckpr4lensing} likelihood, which combines the Planck PR4 lensing potential reconstruction, to account for gravitational lensing effects and enhance the constraints on large-scale structure~\cite{carron2022cmb}.

		\item \textbf{SN Ia}: We make use of three distinct SN Ia datasets to constrain  cosmological parameters. 
		
		The PantheonPlus dataset~\cite{brout2022PantheonPlus}, consisting of 1550 spectroscopically confirmed SN Ia over the redshift range $ 0.001 < z < 2.26 $, is one of the most comprehensive datasets currently available. The likelihood function used in this study is derived from the public release in~\cite{brout2022PantheonPlus}, which incorporates both statistical and systematic covariances. A lower redshift cut of $ z > 0.01 $ is imposed to reduce the impact of peculiar velocities on the Hubble diagram, as described in~\cite{peterson2022PantheonPlus}. This ensures that the distance-redshift relation remains accurate by mitigating local motion effects. 

		The Union3 dataset~\cite{rubin2023union}, which includes a total of 2087 SN Ia, shares 1363 SN Ia in common with PantheonPlus, allowing for direct comparisons and cross-validation. Union3 features a distinct approach to handling systematic uncertainties, employing \textit{Bayesian Hierarchical Modelling} to account for observational biases and measurement errors.
		
		Lastly, the DESY5 dataset~\cite{abbott2024dark}, which is the sample of SN Ia discovered during the full five years of the Dark Energy Survey (DES) Supernova Program, presents a new, homogeneously selected sample of 1635 photometrically classified SN Ia, covering the redshift range $ 0.1 < z < 1.3 $. Notably, this dataset includes a significant number of SN Ia at higher redshifts $z>0.5$ compared to the PantheonPlus sample, which allows for improved constraints on the expansion history of the universe during epochs when dark energy became more prominent. Additionally, it includes 194 low-redshift SN Ia in the range $ 0.025 < z < 0.1 $. The homogeneity of the DESY5 selection helps to minimize potential biases, ensuring reliable cosmological analysis.
	\end{itemize}

	\section{Results and Discussions}
	\label{sec3} 
	
	\begin{table}[htbp]
		\centering
		\begin{tabular}{c|c|c|c|c|c}
			\hline
			& DESI & DESI+CMB &   DESI+CMB+DESY5 & DESI+CMB+Union3 &  DESI+CMB+PantheonPlus \\
			\hline
			$w_{0}$ & $-0.54_{-0.21}^{+0.38}$ & $-0.46_{-0.22}^{+0.33}$ & $-0.73 \pm 0.07$ & $-0.66 \pm 0.1$ & $-0.83 \pm 0.06$\\
			$w_{a}$ & $-1.66_{-1.3}^{+0.43}$ & $-1.78_{-0.98}^{+0.52}$ & $-1.01 \pm 0.30$ & $-1.25_{-0.34}^{+0.41}$ & $-0.73_{-0.25}^{+0.29}$\\
			$\Omega_{m0}$ & $0.345_{-0.025}^{+0.044}$ & $0.343_{-0.028}^{+0.032}$ & $0.3160 \pm 0.0067$ & $0.3229 \pm 0.0096$ & $0.3084 \pm 0.0068$\\
			\hline
		\end{tabular}
		\caption{Mean values and $68 \%$ uncertainties of $w_{0}$, $w_{a}$ and $\Omega_{m0}$.}
		\label{tab1}
	\end{table}
	
	\begin{figure}[htp!]
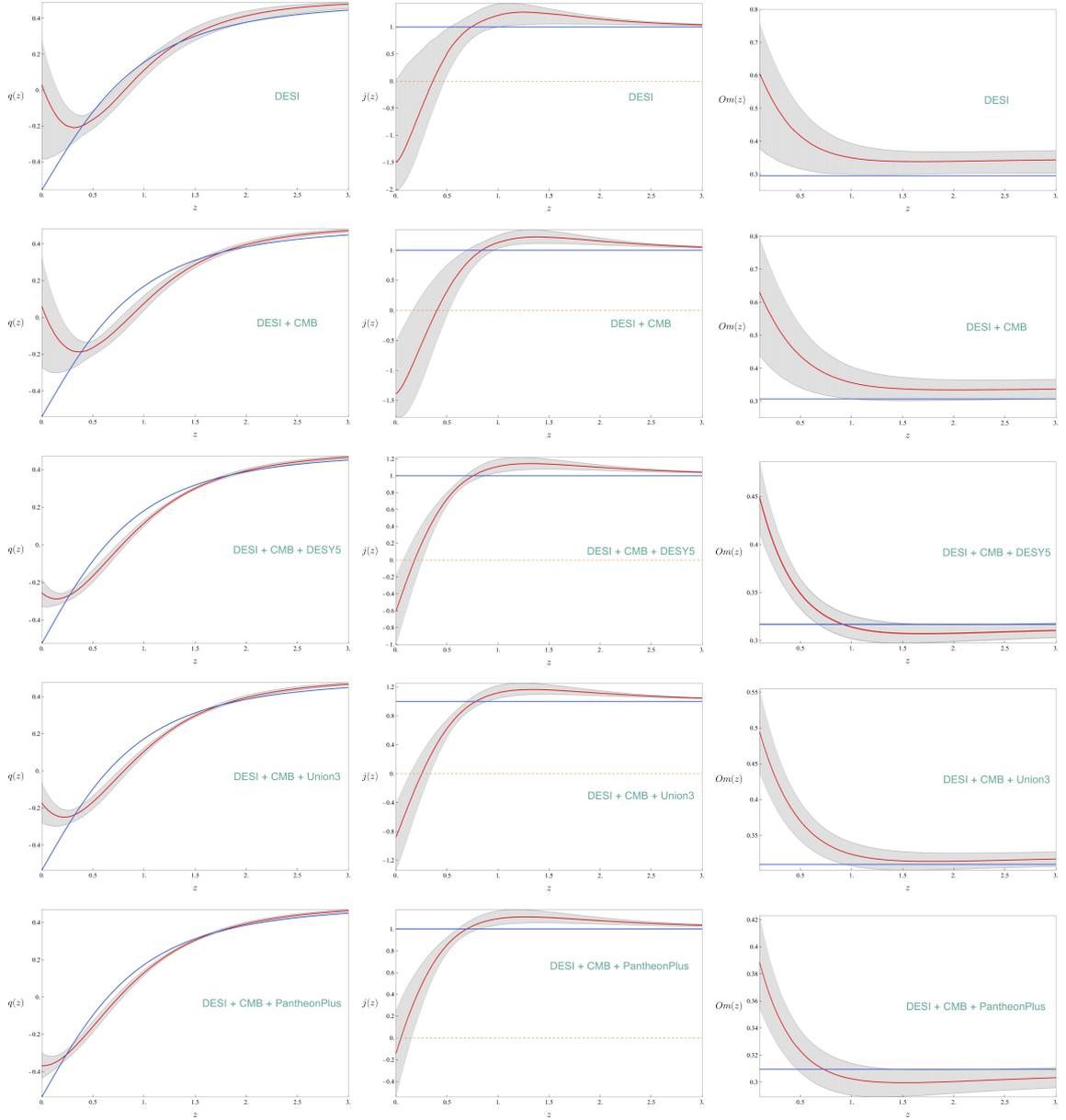

		\centering
	\includegraphics[width=0.3\textwidth]{Fig1a.pdf}	
	\includegraphics[width=0.3\textwidth]{Fig1b.pdf}	
	\includegraphics[width=0.3\textwidth]{Fig1c.pdf}	
	\caption{Evolutions of $q(z)$, $j(z)$ and $Om(z)$ in redshift region of $z < 3$, inferred from DESI BAO, DESI BAO + CMB, DESI BAO + CMB + DESY5, DESI BAO + CMB + Union3, and DESI BAO + CMB + PantheonPlus. The red lines represent the median value for the  $w_0 w_a $CDM model, and the gray shaded regions indicate the $1\sigma$ uncertainties. The blue lines show the best-fitting results of  the  $\Lambda$CDM model.}
		\label{fig1}
	\end{figure}

	\begin{figure}[htp!]
		\centering
		\includegraphics[width=0.45\textwidth]{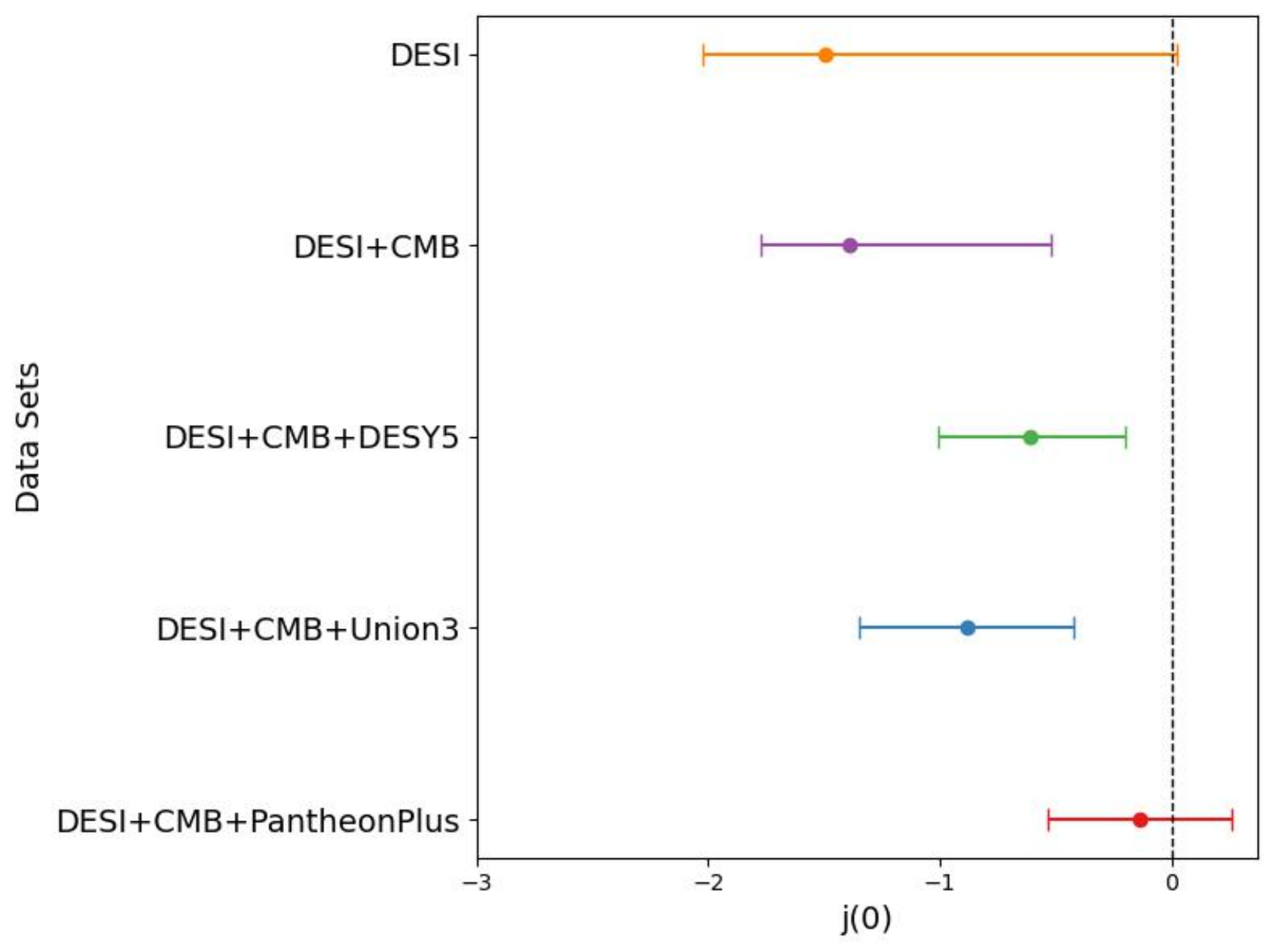}
		\includegraphics[width=0.45\textwidth]{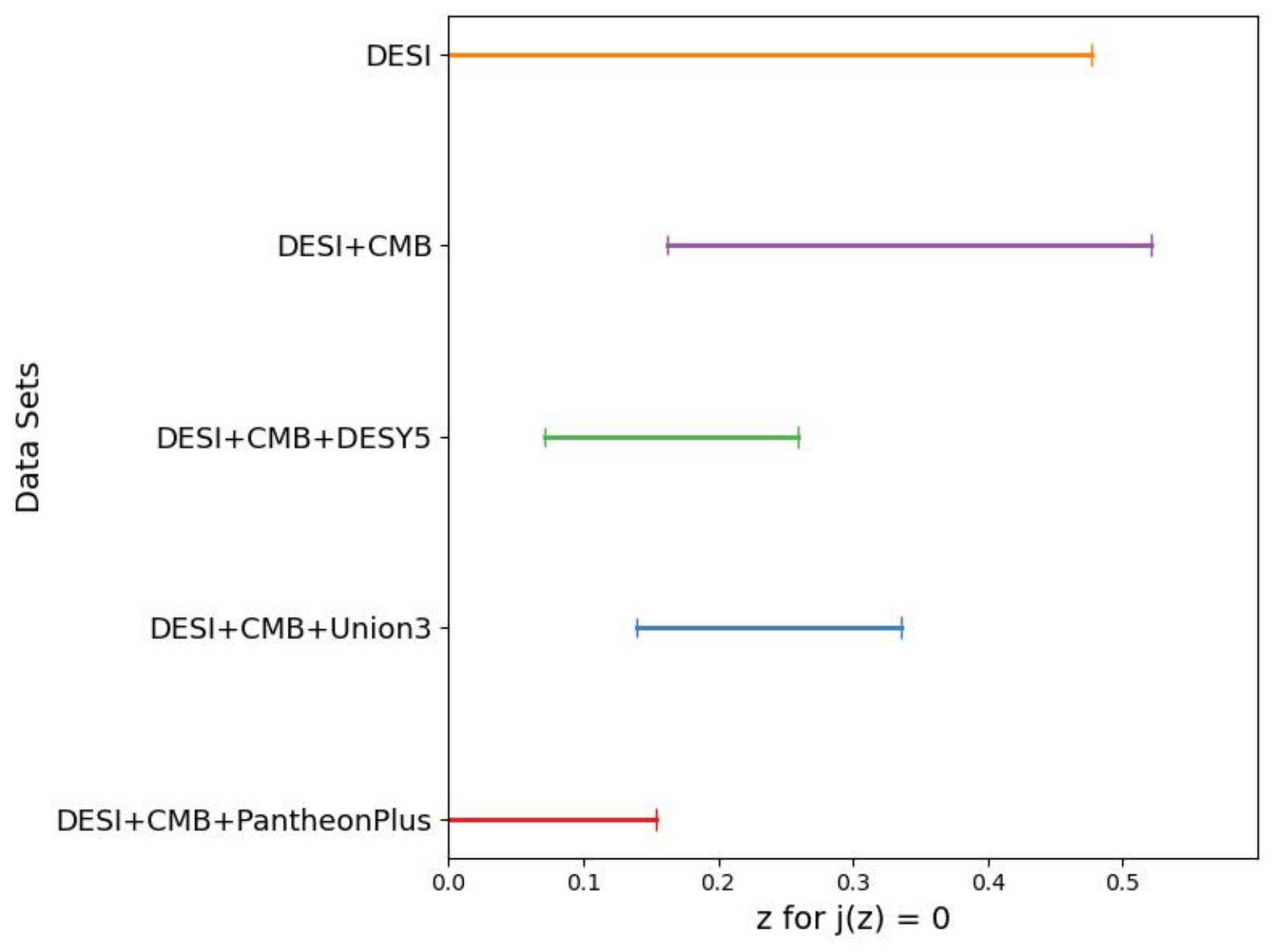}
		\caption{ Constraints of $j(0)$ (left) and $z$ from $j(z) = 0$ (right)  with 68$\%$ CL from different combinations of observational data.}
		\label{fig2}
	\end{figure}

Using  the observed data, we can obtain the constraints on the parameters of the $w_0 w_a$CDM model. In Tab.~\ref{tab1} we summarize the mean values with  $68 \%$ uncertainties of $w_{0}$, $w_{a}$ and $\Omega_{m0}$,  which are used to derive the evolutions of   $q(z)$, $j(z)$  and $Om(z)$. 
Their  evolutionary behaviors  are shown  in Fig.~\ref{fig1}, where the left, middle and right columns  represent $q(z) $, $j(z)$ and $ Om(z) $, respectively, and  rows from top to bottom  are   obtained from DESI BAO,  DESI BAO + CMB, DESI BAO + CMB + DESY5, DESI BAO + CMB + Union3, and  DESI BAO + CMB + PantheonPlus, respectively. The best-fitting  results of the $\Lambda$CDM model are drawn  with blue lines  for comparison. 
	
The DESI BAO data support that   the deceleration parameter may have reached its minimum at redshift about 0.4 and then increases  with the decrease of redshift.  The $q(z)$ evolution deviates from that of the $\Lambda$CDM model in the low redshift region. After adding the Planck CMB data, one can see that the evolutionary characteristics  of $q(z)$ found from DESI BAO become clearer and the deviation from the $\Lambda$CDM model is also more serious.  	These results are different from what have been obtained in \cite{shafieloo2009cosmic} where  the decrease of cosmic acceleration has been found  to disappear at late times,  and the results are perfectly consistent with the $\Lambda$CDM model once the CMB data is included. When the Union3 and DESY5 SN Ia data samples are further included, we find that the  cosmic acceleration still decreases at late times, but the minimum of $q(z)$ seems to occur at a lower redshift compared with the results from DESI BAO and  DESI BAO  + CMB.   However,  DESI BAO  + CMB + PantheonPlus seems to allow an increase of cosmic acceleration at late times although the evolution of $q(z)$ remains different  from the prediction of the $\Lambda$CDM model.  Since all datasets yield deceleration parameter values higher than those predicted by the $\Lambda$CDM model at late times, observations suggest a present-day equation of state for dark energy greater than $-1$.   The median  values of $q(0)$ with $1\sigma$ uncertainties are summarized in Tab.~\ref{tab2}. From this table, one can see that once the SN Ia data are included observations support a present  accelerating cosmic expansion  at more than $1\sigma$ CL. 

From the middle column of Fig.~\ref{fig1}, where the evolution of the jerk parameter is plotted, one can see that  a significant deviation of the $j(z)$ evolution from the $\Lambda$CDM model.  At late times, DESI BAO favors $j(z)<0$ at the margin of $1\sigma$. 
Notably, DESI BAO + CMB, DESI BAO + CMB + DESY5, and DESI BAO + CMB + Union3 support  that the $j(z)$ parameter passes through the $j=0$ line in the low redshift region at more than $1\sigma$ CL.
In contrast, DESI BAO + CMB + PantheonPlus still allows $j(z)>0$ within $1\sigma$ CL, although the likelihood of $j(0)<0$ is slightly larger than $j(0)>0$.  The median  values  of $j(0)$ are shown in Tab.~\ref{tab2}.
To  clearly  show $j(0)$ and the redshift  at which  $j(z)$ passes through $j(z)=0$, we plot Fig.~\ref{fig2}. The left panel of Fig.~\ref{fig2} displays $j(0)$.  This panel and Tab. \ref{tab2}   show that, except for DESI BAO + CMB + PantheonPlus, the preference for $j(0)<0$ exceeds  $1\sigma$ in all cases.   DESI BAO + CMB + PantheonPlus    yields $j(0) = -0.14 \pm 0.4$, making it inconclusive regarding the evolution of cosmic acceleration.  However,  as the best-fitting value of $j(0)$ is negative, this dataset still suggests a greater than  $50\%$  probability that cosmic acceleration is slowing down, an inference that cannot be directly drawn from the  $q(z)$ evolution alone.  
From the right panel of Fig.~\ref{fig2}, we can see that DESI BAO + CMB, DESI BAO + CMB + DESY5 and DESI BAO + CMB + Union3 give the redshifts at which  $j(z)=0$ occur are in the ranges $[0.16 \sim 0.52]$, $[0.07 \sim 0.26]$, and $[0.14 \sim 0.34]$, respectively, at $1\sigma$ CL. These redshifts are larger than zero, indicating  that cosmic acceleration  has already peaked and entered a decelerating phase.

\begin{table}[htbp]
	\centering
	\begin{tabular}{c|c|c|c|c|c}
		\hline
		  & DESI & DESI+CMB &   DESI+CMB+DESY5 & DESI+CMB+Union3 &  DESI+CMB+PantheonPlus \\
		\hline
		$q(0)$ & $0.03_{-0.41}^{+0.25}$ & $0.06_{-0.33}^{+0.27}$ & $-0.26 \pm 0.07$ & $-0.17 \pm 0.11$ & $-0.37 \pm 0.07$   \\
		$j(0)$ & $-1.5_{-0.53}^{+1.52}$ & $-1.39_{-0.38}^{+0.87}$ & $-0.6 \pm 0.4$ & $-0.88 \pm 0.46$ & $-0.14 \pm 0.4$   \\
		\hline
	\end{tabular}
	\caption{Median values and $68 \%$ uncertainties  of $q(0)$ and $j(0)$.}
	\label{tab2}
\end{table}

\begin{figure}[htp!]  
	\centering
	\includegraphics[width=0.7\textwidth]{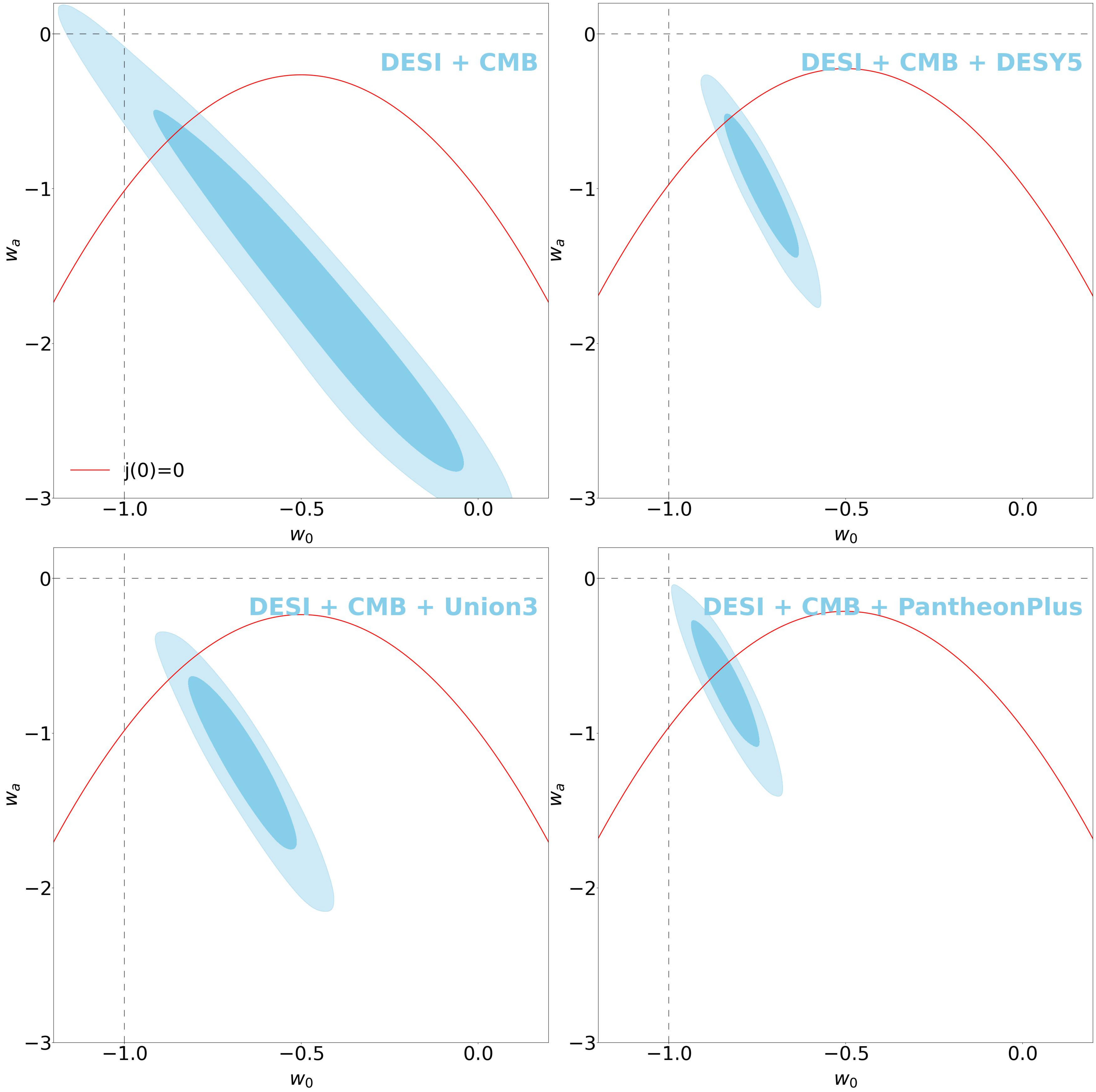}	
	\caption{ Constraints on $w_0$ and $w_a$ from different combinations of datasets: DESI BAO + CMB, DESI BAO + CMB + DESY5, DESI BAO + CMB + Union3, and DESI BAO + CMB + PantheonPlus. The dark and  light regions correspond to the $1\sigma$ and $2\sigma$ CLs, respectively. The red line represents Eq. (\ref{Eq6}) with $\Omega_{m0}$ set to the best-fit value for each dataset, which are  $\Omega_{m0}=0.343, \ 0.316, \ 0.323$, and  $0.308$ for DESI BAO + CMB, DESI BAO + CMB + DESY5, DESI BAO + CMB + Union3, and DESI BAO + CMB + PantheonPlus, respectively. The regions below this red line mean that parameters $w_0$ and $w_a$ lead to $j(0)<0$.}
	\label{fig3}
\end{figure}

 We find that when $w_0$ and $w_a$ satisfy the condition
\begin{eqnarray}\label{Eq6}
w_a< -3w_0(w_0+1)-\frac{2}{3(1-\Omega_{m0})}
\end{eqnarray}
the value of $j(0)$ becomes $j(0)<0$. To show clearly the condition in Eq.~(\ref{Eq6}) in the $w_0-w_a$ space, we plot the constraints on the CPL  parametrization in Fig.~\ref{fig3} and use the red line to indicate this condition with $\Omega_{m0}$ set to the best-fit value for each dataset. We find that in the cases of DESI BAO + CMB, DESI BAO + CMB + DESY5 and DESI BAO + CMB + Union3  only a very small parameter space in the $w_0-w_a$  plane supports $j(0)>0$.   The parameter space in the $w_0-w_a$  plane allowing $j(0)<0$ is slightly larger than the one for $j(0)>0$ when DESI BAO + CMB + PantheonPlus is used.

To verify these findings, we apply the $Om(z)$ diagnostic. The results are shown in the right column of Fig.~\ref{fig1}. One can see that at late times  the $Om(z)$  evolutions deviate apparently from the prediction of the $\Lambda$CDM model  and prefer a value of the equation of state of dark energy  larger than $-1$.  At low redshift region ($z\ll1$), we have  $w \approx [Om(z)-1]/(1-\Omega_{m0})$). However,  the right column of Fig.~\ref{fig1} represents clearly that $Om(z)$ evolves with redshift in the $z\ll1$ region, which means that  the $Om(z)$ diagnostic may favor a dynamical dark energy.  These results are well consistent with those obtained  from the left and middle columns of Fig.~\ref{fig1} and Fig.~\ref{fig3}.

	\section{conclusions}
	\label{sec4}

In this paper, we revisit the evolution of cosmic acceleration in a spatially flat $w_0w_a$CDM universe,  in which the equation of state of dark energy is described by the CPL parametrization,  by using  DESI BAO, Planck  CMB, and several SN Ia datasets, including PantheonPlus, Union3, and DESY5. In addition to analyzing the evolution of the deceleration parameter, we also consider the jerk parameter, which characterizes the variation of  cosmic acceleration with redshift, and further check our findings  using  the $Om(z)$ diagnostic. We find that within the framework of  the $w_0w_a$CDM model  the evolutions of  the deceleration and jerk parameters deviate significantly from the predictions of the $\Lambda$CDM model, while  the $Om(z)$ diagnostic prefers  dynamical dark energy. The datasets  DESI BAO, DESI BAO + CMB, DESI BAO + CMB + Union3, and DESI BAO + CMB + DESY5 all provide strong evidence for a slowing down of cosmic acceleration at late times, since  $j(0)$ falls below zero at more than $1\sigma$ CL.  In contrast,  DESI BAO + CMB + PantheonPlus only slightly supports a decreasing acceleration, as it allows $j(0)=0$  within  $1\sigma$ CL.  Moreover,  DESI BAO + CMB, DESI BAO + CMB + DESY5 and DESI BAO + CMB + Union3 all support that   cosmic  acceleration has  already entered   a decreasing phase, as  they yield the redshifts for $j(z)=0$ that are larger than zero  at more than $1\sigma$ CL.  These results contrast with the findings of \cite{shafieloo2009cosmic},  where the decline in cosmic acceleration observed from SN Ia + BAO disappeared when CMB data from the WMAP survey was included in the analysis, resulting in consistency with the $\Lambda$CDM model . Thus, our results suggest that  in  the $w_0w_a$CDM model most of the latest observational data favor the scenario in which cosmic acceleration has already peaked and is now entering a phase of gradual decline. Finally, we note that shortly after the completion of this manuscript, DESI released its Data Release 2 (DR2)~\cite{Karim2025}.  We have verified that incorporating the DESI DR2 data does not alter the main results presented in this work.  

\begin{acknowledgments}
This work was supported in part by the NSFC under Grant  Nos. 12275080 and 12075084,  and the Innovative Research Group of Hunan Province under Grant No.~2024JJ1006.

\end{acknowledgments}


\begin{thebibliography}{99}
	
\bibitem{riess1998observational}
A. G. Riess, A. V. Filippenko, P. Challis,  et al.,
\href{https://doi.org/10.1086/300499}
{Astrophys. J. \textbf{116}, 1009 (1998)}.

\bibitem{perlmutter1999measurements}
S. Perlmutter, G. Aldering, G. Goldhabe,  et al.,
\href{https://doi.org/10.1086/307221}
{Astron. J. \textbf{517}, 565 (1999)}.

\bibitem{Daniel2005}
D. J. Eisenstein, I. Zehavi, D. W. Hogg,  et al.,
\href{https://doi.org/10.1086/466512}
{Astrophys. J. \textbf{633}, 560 (2005)}.

\bibitem{hinshaw2013nine}
G. Hinshaw,  D. Larson, E. Komatsu, et al.,
\href{https://doi.org/10.1088/0067-0049/208/2/19}
{Astrophys. J. Suppl. \textbf{208}, 19 (2013)}.

\bibitem{aghanim2020planck}
N. Aghanim, Y. Akrami, M. Ashdown, et al.,
\href{https://doi.org/10.1051/0004-6361/201833910}{Astron. Astrophys. \textbf{641}, A6 (2020)}.

\bibitem{shafieloo2009cosmic}
A. shafieloo,   V. Sahni, and A. A. Starobinsky, 
\href{https://doi.org/10.1103/PhysRevD.80.101301}
{Phys. Rev. D \textbf{80}, 101301 (2009)}.

\bibitem{hicken2009improved}
M. Hicken, W. M. Wood-Vasey, S. Blondin, et al.,
\href{https://doi.org/10.1088/0004-637X/700/2/1097}
{Astrophys. J. \textbf{700}, 1097 (2009)}.

\bibitem{hicken2009cfa3}
M. Hicken, P. Challis, S. Jha, et al.,
\href{https://doi.org/10.1088/0004-637X/700/1/331}
{Astrophys. J. \textbf{700}, 331 (2009)}.

\bibitem{percival2010baryon}
W. J. Percival,   B. A. Reid,   D. J. Eisenstein, et al.,
\href{https://doi.org/10.1111/j.1365-2966.2009.15812.x}
{Mon. Not. Roy. Astron. Soc. \textbf{401}, 2148 (2010)}.

\bibitem{WMAP:2008lyn}
E. Komatsu, J. Dunkley, M.R. Nolta, et al., \href{https://doi.org/10.1088/0067-0049/180/2/330}{Astrophys. J. Suppl. {\bf 180}, 330 (2009)}.
  
\bibitem{Fabris2010}
J. C. Fabris, B. Fraga, N. P. Neto and W. Zimdahl,
\href{https://doi.org/10.1088/1475-7516/2010/04/008}
{J. Cosmol. Astropart. Phys. \textbf{04} (2010) 008 }.

\bibitem{Gong2010}
Y. Gong, B. Wang, and R. Cai, \href{https://doi.org/10.1088/1475-7516/2010/04/019}{J. Cosmol. Astropart. Phys. \textbf{04} (2010) 019}.

\bibitem{Guimaraes2011}
A. C. C. Guimar\~aes and J. A. S. Lima,
\href{https://doi.org/10.1088/0264-9381/28/12/125026}
{Class. Quantum Grav. \textbf{28}, 125026 (2011)}.

\bibitem{Li2010}
Z. Li, P. Wu and H. Yu, 
\href{https://doi.org/10.1088/1475-7516/2010/11/031}{J. Cosmol. Astropart. Phys. {\bf 11} (2010) 031}.


\bibitem{Li2011}
Z. Li, P. Wu and H. Yu, 
\href{https://doi.org/10.1016/j.physletb.2010.10.044}{Phys. Lett. B {\bf 695}, 1 (2011)}.

\bibitem{Li20112}
X. D. Li, S. Li, S. Wang, W. S. Zhang, Q. G. Huang and M. Li,
\href{https://doi.org/10.1088/1475-7516/2011/07/011}
{J. Cosmol. Astropart. Phys. \textbf{07} (2011) 011}.

\bibitem{Cai2011}
R. G. Cai and Z. L. Tuo,
\href{https://doi.org/10.1016/j.physletb.2011.11.021}
{Phys. Lett. B \textbf{706}, 116 (2011)}.

\bibitem{Bolotin2012}
Y. L. Bolotin, O. A. Lemets,  and D. A. Yerokhin, \href{https://doi.org/10.3367/UFNe.0182.201209c.0941}{Phys. Usp. {\bf 55}, 876 (2012)}.

\bibitem{Vargas2012}
C. Z. Vargas, W. S. Hipolito-Ricaldi, and W. Zimdahl, \href{https://doi.org/10.1088/1475-7516/2012/04/032}{J. Cosmol. Astropart. Phys. \textbf{04} (2012) 032}.

\bibitem{Cardenas2012}
V. H. Cardenas and M. Rivera, \href{https://doi.org/10.1016/j.physletb.2012.03.004}{Phys. Lett. B {\bf 710}, 251 (2012)}.

\bibitem{Lin2013}
J. Lin, P. Wu and H. Yu, \href{https://doi.org/10.1103/PhysRevD.87.043502}{Phys. Rev. D {\bf 87}, 043502 (2013)}.

\bibitem{Magana2014}
J. Maga\~na, V. H. Cadenas and V. Motta,
\href{https://doi.org/10.1088/1475-7516/2014/10/017}
{J. Cosmol. Astropart. Phys. \textbf{10} (2014) 017}.

\bibitem{Shahalam2015}
M. Shahalam, S. Sami and A. Agarwal, \href{https://doi.org/10.1093/mnras/stv083}{Mon. Not. Roy. Astron. Soc. {\bf 448}, 2948 (2015)}.

\bibitem{Bernal2017}
C. Bernal, V. H. C\'ardenas, V. Motta,
\href{https://doi.org/10.1016/j.physletb.2016.12.008}
{Phys. Lett. B \textbf{765}, 163 (2017)}.

\bibitem{Zhang2018}
M. J. Zhang, J. Q. Xia,
\href{https://doi.org/10.1016/j.nuclphysb.2018.02.020}
{Nucl. Phys. B \textbf{929}, 438 (2018)}.

\bibitem{Velten2018}
H. Velten, S. Gomes, and V. C. Busti, \href{https://doi.org/10.1103/PhysRevD.97.083516}{Phys. Rev. D {\bf 97},   083516 (2018)}.

\bibitem{Haridasu2018}
B. S. Haridasu, V. V. Lukovic, M. Moresco, and N. Vittorio, \href{https://doi.org/10.1088/1475-7516/2018/10/015}{J. Cosmol. Astropart. Phys. \textbf{10} (2018) 015}.

\bibitem{Fortunato2024}
J. A. S. Fortunato, W. S. Hipolito-Ricaldi, N. Videla, and J. R. Villanueva, \href{https://doi.org/10.1140/epjc/s10052-025-13996-3}{Eur. Phys. J. C {\bf 85}, 274 (2025)}.


\bibitem{Wang2025}
D. Wang, K. Bamba,
\href{https://doi.org/10.48550/arXiv.2506.23029}{arXiv:2506.23029}

\bibitem{Adame2024a}
 A. G. Adame, J. Aguilar, S. Ahlen, et. al., 
 \href{https://doi.org/10.3847/1538-3881/ad3217}{Astron. J. {\bf 168}, 58 (2024)}.

\bibitem{Adame2024desi}
A. G. Adame, J. Aguilar, S. Ahlen, et al.,
\href{https://doi.org/10.1088/1475-7516/2025/02/021}
{J. Cosmol. Astropart. Phys. \textbf{02} (2025) 021}.

\bibitem{brout2022PantheonPlus}
D. Brout, D. Scolnic, B. Popovic,  et al.,
\href{https://doi.org/10.3847/1538-4357/ac8e04}
{Astrophys. J. \textbf{938}, 110  (2022)}.

\bibitem{rubin2023union}
D. Rubin,  G. Aldering, M. Betoule, et al.,
\href{https://doi.org/10.48550/arXiv.2311.12098}
{arXiv: 2311.12098}.

\bibitem{abbott2024dark}
T. M. C. Abbott,  M. Acevedo, M. Aguena, et al.,
\href{https://doi.org/10.48550/arXiv.2401.02929}
{ Astrophys. J. Lett. {\bf 973}, L14 (2024)}.

\bibitem{Calderon2024}
R. Calderon et.al,
\href{https://doi.org/10.1088/1475-7516/2024/10/048}{J. Cosmol. Astropart. Phys. \textbf{10} (2024) 048}.

\bibitem{Raamsdonk2024}
M. V. Raamsdonk and C. Waddell, \href{https://doi.org/10.1088/1475-7516/2024/06/047}{J. Cosmol. Astropart. Phys. \textbf{06} (2024) 047}

\bibitem{Tada2024}
Y. Tada and T. Terada, \href{https://doi.org/10.1103/PhysRevD.109.L121305}{Phys. Rev. D {\bf 109}, L121305  (2024)}.

\bibitem{Yin2024}
W. Yin, \href{https://doi.org/10.1007/JHEP05(2024)327}{JHEP 05 (2024) 327 }.

\bibitem{Luongo2024}
O. Luongo and M. Muccino, \href{https://doi.org/10.1051/0004-6361/202450512}{Astron. Astrophys. {\bf 690},  A40 (2024)}.


\bibitem{Cortes2024}
M. Cortes and A. R. Liddle, \href{https://doi.org/10.1088/1475-7516/2024/12/007}{J. Cosmol. Astropart. Phys. {\bf 12} (2024) 007}.

\bibitem{DWang2024}
D. Wang, \href{https://arxiv.org/abs/2404.06796}{arXiv: 2404.06796}.

\bibitem{Colgain2024a}
E. O. Colgain, M. G. Dainotti, S. Capozziello, et. al., 
\href{https://arxiv.org/abs/2404.08633}{arXiv:2404.08633}.

\bibitem{Berghaus2024}
K. V. Berghaus, J. A. Kable, and V. Miranda, \href{https://doi.org/10.1103/PhysRevD.110.103524}{Phys. Rev. D {\bf 100}, 103524 (2024)}.

\bibitem{Giare2024a}W. Giare, M. A. Sabogal, R. C. Nunes, and E. Di Valentin, \href{https://doi.org/10.1103/PhysRevLett.133.251003}{Phys. Rev. Lett. {\bf 133}, 251003  (2024)}.

\bibitem{Wang20242}
H. Wang and Y. Piao, \href{https://arxiv.org/abs/2404.18579}{arXiv: 2404.18579}.

\bibitem{Yang2024}Y. Yang, X. Ren, Q. Wang, Z. Lu, D. Zhang, Y. Cai,  and E. N. Saridakis, \href{
https://doi.org/10.1016/j.scib.2024.07.029}{Science Bulletin {\bf 69},  2698  (2024)}.

\bibitem{Park2024}
C. Park, J. de Cruz Perez, and B. Ratra, \href{https://doi.org/10.1103/PhysRevD.110.123533}{Phys. Rev. D {\bf 110},   123533  (2024)}.

\bibitem{Wang20243}
Z. Wang, S. Lin, Z.  Ding,  and B. Hu, \href{https://doi.org/10.1093/mnras/stae2309}{Mon. Not. Roy. Astron. Soc. {\bf 534},   3869 (2024)}.

\bibitem{Huang2024}
Z. Huang, J. Liu, J. Mo, et al., \href{https://doi.org/10.1103/PhysRevD.110.123512}{Phys. Rev. D {\bf 110},     123512 (2024)}.

\bibitem{Dinda2024}
B. R. Dinda, \href{https://doi.org/10.1088/1475-7516/2024/09/062}{J. Cosmol. Astropart. Phys. \textbf{09} (2024) 062}.

\bibitem{Bousis2024}
D. Bousis and L. Perivolaropoulos, \href{https://doi.org/10.1103/PhysRevD.110.103546}{Phys. Rev. D {\bf 110},  103546  (2024)}.

\bibitem{Lodha2025}
K. Lodha, A. Shafieloo, R. Calderon, et al., \href{https://doi.org/10.1103/PhysRevD.111.023532}{Phys. Rev. D {\bf 111}, 023532 (2025)}.

\bibitem{Ramadan2024}
O. F. Ramadan and J.  Sakstein, \href{https://doi.org/10.1103/PhysRevD.110.L041303}{Phys. Rev. D {\bf 110},  L041303  (2024)}.

\bibitem{Roy2024}
N. Roy, \href{https://doi.org/10.1016/j.dark.2025.101912}{Phys. Dark Universe {\bf 48}, 101912 (2025)}. 


\bibitem{Wang20244}
H. Wang, Z. Peng, Y. Piao, \href{https://arxiv.org/abs/2406.03395}{arXiv: 2406.0339}.

\bibitem{Gialamas2025}
I. D. Gialamas, G. Hutsi, K. Kannike, A. Racioppi,  and M. Raidal, \href{https://doi.org/10.1103/PhysRevD.111.043540}{Phys. Rev. D {\bf 111}, 043540 (2025)}.

\bibitem{Chudaykin2024}
A. Chudaykin and  M. Kunz, \href{https://doi.org/10.1103/PhysRevD.110.123524}{Phys. Rev. D {\bf 110},  123524  (2024)}.

\bibitem{Orchard2024}
L. Orchard and V. H. Cardenas, \href{https://doi.org/10.1016/j.dark.2024.101678}{Phys. Dark Univ. {\bf 46} 101678  (2024) }.

\bibitem{Li2024}
T. Li, P. J. Wu, G. Du, S. Jin, H. Li, J. Zhang, and X. Zhang, \href{https://doi.org/10.3847/1538-4357/ad87f0}{Astrophys. J. {\bf 976}, 1 (2024)}.

\bibitem{Du2024}
G. Du, P. J. Wu, T. Li, and  X. Zhang, \href{https://doi.org/10.1140/epjc/s10052-025-14094-0}{Eur. Phys. J. C {\bf 85}, 392 (2025)}.


\bibitem{Giare2024}
W. Giare, M.  Najafi, S. Pan, E. Di Valentino, J. T. Firouzjaee, \href{https://doi.org/10.1088/1475-7516/2024/10/035}{J. Cosmol. Astropart. Phys. {\bf 10} (2024) 035}.


\bibitem{Dinda2025}
B. R. Dinda and R. Maarten, \href{https://doi.org/10.1088/1475-7516/2025/01/120}{J. Cosmol. Astropart. Phys. {\bf 01} (2025) 120}.

\bibitem{Pang2024}
Y. Pang, X. Zhang and Q. Huang, \href{https://arxiv.org/abs/2408.14787}{arXiv: 2408.14787}.

\bibitem{David2024}
D. Shlivko and P. J. Steinhardt,
\href{https://doi.org/10.1016/j.physletb.2024.138826}{Phys. Lett. B {\bf 855}, 138826 (2024)}.

\bibitem{Wolf2024}
W. J. Wolf, C. G. Garc\'{i}a, D. J. Bartlett and P. G. Ferreira,
\href{https://doi.org/10.1103/PhysRevD.110.083528}{Phys. Rev. D \textbf{110}, 083528 (2024)}.

\bibitem{Almada2024}
A. H. Almada, M.L. M. Martínez, M. A. G. Aspeitia and V. Motta,
\href{https://doi.org/10.1016/j.dark.2024.101668}{Phys. Dark Universe {\bf 46}, 101668 (2024)}.

\bibitem{Li20242}
T. Li, Y. Li, G. Du, P.  Wu, L. Feng, J. Zhang, and X. Zhang, \href{https://arxiv.org/abs/2411.08639}{arXiv: 2411.0863}.

\bibitem{Colgain2024}
E. O. Colgain and M.M. Sheikh-Jabbari, \href{https://arxiv.org/abs/2412.12905}{arXiv: 2412.12905}.

\bibitem{Zheng2024}
J. Zheng, D. Qiang, and Z. You, \href{https://arxiv.org/abs/2412.04830}{arXiv: 2412.04830}.

\bibitem{Gao2024}
Q. Gao, Z. Peng, S. Gao,  and Y. Gong, \href{https://doi.org/10.3390/universe11010010}{Universe {\bf 11}, 10 (2025)}.

\bibitem{Notari2024}
A. Notari, M. Redi, and  A. Tesi, \href{https://arxiv.org/abs/2411.11685}{arXiv: 2411.11685}.

\bibitem{Odintsov2024}
S. D. Odintsov, D. S. C. G\'{o}mez, G. S. Sharov, \href{https://doi.org/10.48550/arXiv.2412.09409}{arXiv:2412.09409}.

\bibitem{Odintsov2025a}
S. D. Odintsov, V. K. Oikonomou, G. S. Sharov, \href{https://doi.org/10.48550/arXiv.2503.17946}{arXiv:2503.17946}.

\bibitem{Odintsov2025b}
S. D. Odintsov, V. K. Oikonomou, G. S. Sharov, \href{https://doi.org/10.48550/arXiv.2506.02245}{arXiv:2506.02245}.

\bibitem{Marcus2025}
M. H{\"o}g{\aa}s, E. M\"{o}rtsell, \href{https://doi.org/10.48550/arXiv.2507.03743}{arXiv:2507.03743}.

\bibitem{Mishra2025}
S. S. Mishra, W. L. Matthewson, V. Sahni, A. Shafieloo, Y. Shtanov \href{https://doi.org/10.48550/arXiv.2507.07193}{arXiv:2507.07193}.

\bibitem{Sousa2025}
A. Sousa-Neto, C. Bengaly, J. E. Gonzalez, and J. Alcaniz, 
\href{https://arxiv.org/abs/2502.10506}{arXiv: 2502.10506}

\bibitem{Huang2025}
L. Huang, R. Cai, and S. Wang, \href{https://arxiv.org/abs/2502.04212}{arXiv: 2502.04212}.

\bibitem{Li2025}
T. Li, G. Du, Y. Li, P. Wu, S. Jin, J. Zhang, and X. Zhang, \href{https://arxiv.org/abs/2501.07361}{arXiv: 2501.07361}.

\bibitem{Giare2025}
W. Giar\`{e}, T. Mahassen, E. D. Valentino and S. Pan,
\href{https://doi.org/10.1016/j.dark.2025.101906}{Phys. Dark Universe {\bf 48}, 101906 (2025)}.

\bibitem{Efstathiou2025}
G. Efstathiou,
\href{https://doi.org/10.1093/mnras/staf301}{Mon. Not. Roy. Astron. Soc. {\bf 538}, 875–882 (2025)}.

\bibitem{Liu2025}
Y. Liu, Y. Zhang,  H. Yu, and P. Wu, \href{https://inspirehep.net/literature/2929371}{arXiv:  2506.03536}.

\bibitem{Carloni2025a}
Y. Carloni, O. Luongo, and M. Muccino,
\href{https://doi.org/10.1103/PhysRevD.111.023512}{Phys. Rev. D \textbf{111}, 023512 (2025)}.

\bibitem{Notari2025}
A. Notari, M. Redi and A. Tesi,
\href{https://doi.org/10.1088/1475-7516/2025/04/048}{J. Cosmol. Astropart. Phys., \textbf{04} (2025) 048}.

\bibitem{Valent2025}
A. G. Valent and J. S. Peracaula,
\href{https://doi.org/10.1016/j.physletb.2025.139391}{Phys. Lett. B {\bf 864}, 139391 (2025)}.

\bibitem{Pang2025}
Y. H. Pang, X. Zhang and Q. G. Huang,
\href{https://doi.org/10.1103/PhysRevD.111.123504}{Phys. Rev. D \textbf{111}, 123504 (2025)}.

\bibitem{Chatrchyan2025}
A. Chatrchyan, F. Niedermann, V. Poulin and M. S. Sloth,
\href{https://doi.org/10.1103/PhysRevD.111.043536}{Phys. Rev. D \textbf{111}, 043536 (2025)}.

\bibitem{Wolf2025}
W. J. Wolf, P. G. Ferreira and C. G. Garc\'{i}a,
\href{https://doi.org/10.1103/PhysRevD.111.L041303}{Phys. Rev. D \textbf{111}, L041303 (2025)}.

\bibitem{Shlivko2025}
D. Shlivko, P. J. Steinhardt and C. L. Steinhardt,
\href{https://doi.org/10.1088/1475-7516/2025/06/054}{J. Cosmol. Astropart. Phys., \textbf{06} (2025) 054}.

\bibitem{LiTN2025}
T. N. Li, Y. H. Li, G. H. Du, et al.,
\href{https://doi.org/10.1140/epjc/s10052-025-14279-7}{Eur. Phys. J. C {\bf 02}, 608 (2025)}.

\bibitem{Taule2025}
P. Taule, M. Marinucci, G. Biselli, M. Pietroni and F. Vernizzi,
\href{https://doi.org/10.1088/1475-7516/2025/03/036}{J. Cosmol. Astropart. Phys., \textbf{03} (2025) 036}.

\bibitem{Qu2025}
F. J. Qu, K. M. Surrao, B. Bolliet, et al.,
\href{https://doi.org/10.1103/xhh6-9v62}{Phys. Rev. D \textbf{111}, 123507 (2025)}.

\bibitem{Tang2025}
X. TZ Tang, D. Brout, T. Karwal, et al.,
\href{https://doi.org/10.3847/2041-8213/adc4da}{Astrophys. J. Lett. {\bf 983}, L27 (2025)}.

\bibitem{Carloni2025}
Y. Carloni and O. Luongo,
\href{https://doi.org/10.1088/1361-6382/adc06e}{Class. Quantum Grav. \textbf{42}, 075014 (2025)}.

\bibitem{Kessler2025}
D. A. Kessler, L. A. Escamilla, S. Pan and E. D. Valentino,
\href{https://doi.org/10.48550/arXiv.2504.00776}{arXiv:2504.00776}.

\bibitem{Chunyu2025}
C. Li, J. Wang, D. Zhang, E. N. Saridakis and Y. F. Cai,
\href{https://doi.org/10.48550/arXiv.2504.07791}{arXiv:2504.07791}.

\bibitem{Chakraborty2025}
A. Chakraborty, P. K. Chanda, S. Das and K. Dutta,
\href{https://doi.org/10.48550/arXiv.2503.10806}{arXiv:2503.10806}.

\bibitem{Wang2025a}
D. Wang,
\href{https://doi.org/10.48550/arXiv.2504.15635}{arXiv:2504.15635}

\bibitem{Braglia2025}
M. Braglia, X. Chen, A. Loeb,
\href{https://doi.org/10.48550/arXiv.2507.13925}{arXiv:2507.13925}

\bibitem{Eoin2025}
E. \'{O}. Colg\'{a}in, S. Pourojaghi, M. M. S. Jabbari, L. Yin,
\href{https://doi.org/10.48550/arXiv.2504.04417}{arXiv:2504.04417}

\bibitem{Gialamas2025a}
I. D. Gialamas, G. H\"{u}tsi, M. Raidal, et al.,
\href{https://doi.org/10.48550/arXiv.2506.21542}{arXiv:2506.21542}

\bibitem{Shafieloo2012}
A. Shafieloo,
\href{https://doi.org/10.1088/1475-7516/2012/08/002}
{J. Cosmol. Astropart. Phys., \textbf{08} (2012) 002}.

\bibitem{alam2003exploring}
U. Alam, V. Sahni, T. D. Saini, A. A. Starobinsky,
\href{https://doi.org/10.1046/j.1365-8711.2003.06871.x}
{Mon. Not. Roy. Astron. Soc. \textbf{344}, 1057 (2003)}.

\bibitem{visser2004jerk}
M. Visser,
\href{https://doi.org/10.1088/0264-9381/21/11/006}
{Class. Quantum Grav. \textbf{21}, 2603 (2004)}.

\bibitem{rapetti2007kinematical}
D. Rapetti, S. W. Allen, M. A. Amin, R. D. Blandford,
\href{https://doi.org/10.1111/j.1365-2966.2006.11419.x}
{Mon. Not. Roy. Astron. Soc. \textbf{375}, 1510 (2007)}.

\bibitem{blandford2004cosmokinetics}
R. D. Blandford, M. Amin, E. A. Baltz, K. Mandel, P. J. Marshall,
\href{https://doi.org/10.48550/arXiv.astro-ph/0408279}
{arXiv:astro-ph/0408279}.

\bibitem{sahni2003statefinder}
V. Sahni, T. D. Saini, A. A. Starobinsky, and U. Alam,
\href{https://doi.org/10.1134/1.1574831}
{Jetp Lett. \textbf{77}, 201 (2003)}.

\bibitem{sahni2008two}
V. Sahni, A. Shafieloo, and A. A. Starobinsky,
\href{https://doi.org/10.1103/PhysRevD.78.103502}
{Phys. Rev. D \textbf{78}, 103502 (2008)}.

\bibitem{zunckel2008consistency}
C. Zunckel, and C. Clarkson,
\href{https://doi.org/10.1103/PhysRevLett.101.181301}
{Phys. Rev. Lett. \textbf{101}, 181301 (2008)}.

\bibitem{chevallier2001accelerating}
M. Chevallier and D. Polarski,
\href{https://doi.org/10.1142/S0218271801000822}
{Int. J.  Mod. Phys.  D \textbf{10}, 213 (2001)}.

\bibitem{linder2003exploring}
E. V. Linder,
\href{https://doi.org/10.1103/PhysRevLett.90.091301}
{Phys. Rev. Lett. \textbf{90}, 091301 (2003)}.

\bibitem{torrado2019cobaya}
E. V. Linder,
{Astrophysics Source Code Library, ascl-1910 (2019)}.

\bibitem{torrado2021cobaya}
J. Torrado, A. Lewis,
\href{https://doi.org/10.1088/1475-7516/2021/05/057}
{J. Cosmol. Astropart. Phys. \textbf{05} (2021) 057}.

\bibitem{lewis2000efficient}
A. Lewis, A. Challinor, and A. Lasenby,
\href{https://doi.org/10.1086/309179}
{Astrophys. J. \textbf{538}, 473 (2000)}.

\bibitem{howlett2012cmb}
C. Howlett, A. Lewis, A. Hall and A. Challinor,
\href{https://doi.org/10.1088/1475-7516/2012/04/027}
{J. Cosmol. Astropart. Phys. \textbf{04} (2012) 027}.

\bibitem{gelman1992inference}
A. Gelman and  D. B. Rubin,
\href{https://doi.org/10.1214/ss/1177011136}
{Statist. Sci., \textbf{7}, 457 (1992)}.

\bibitem{lewis2019getdist}
A. Lewis,
\href{https://doi.org/10.48550/arXiv.1910.13970}
{arXiv:1910.13970}.

\bibitem{adame2024desib}
A. G. Adame, J. Aguilar, S. Ahlen, et al.,
\href{https://doi.org/10.1088/1475-7516/2025/04/012}
{J. Cosmol. Astropart. Phys. {\bf 04}, (2025) 012}.


\bibitem{adame2024desic}
A. G. Adame, J. Aguilar, S. Ahlen,  et al.,
\href{https://doi.org/10.1088/1475-7516/2025/01/124}
{J. Cosmol. Astropart. Phys. \textbf{01} (2025) 124}.

\bibitem{carron2022cmb}
J. Carron, M. Mirmelstein and A. Lewis,
\href{https://doi.org/10.1088/1475-7516/2022/09/039}
{J. Cosmol. Astropart. Phys. \textbf{09} (2022) 039}.

\bibitem{peterson2022PantheonPlus}
E. R. Peterson, W. D'Arcy Kenworthy, D. Scolnic, et al.,
\href{https://doi.org/10.3847/1538-4357/ac4698}
{Astrophys. J. \textbf{938}, 112 (2022)}.

\bibitem{Karim2025}
M. Abdul Karim, J. Aguilar,  S. Ahlen, et al., \href{https://doi.org/10.48550/arXiv.2503.14738}{arXiv:2503.14738}







\end{thebibliography}
\end{document}